# Modeling of the nonlinear flame response of a Bunsen-type flame via multi-layer perceptron

Nilam Tathawadekar [a,d,*], Nguyen Anh Khoa Doan [b,c], Camilo F. Silva [b], Nils Thuerey [d]

[a] *GE Aviation Digital, Freisinger Landstr. 50, Garching 85748, Germany*
[b] *Department of Mechanical Engineering, Technical University of Munich, Boltzmannstr. 15, Garching 85747, Germany*
[c] *Institute for Advanced Study, Technical University of Munich, Lichtenbergstr. 2a, Garching 85748, Germany*
[d] *Department of Informatics, Technical University of Munich, Boltzmannstr. 3, Garching 85748, Germany*

## Abstract

This paper demonstrates the ability of neural networks to reliably learn the nonlinear flame response of a laminar premixed flame, while carrying out only one unsteady CFD simulation. The system is excited with a broadband, low-pass filtered velocity signal that exhibits a uniform distribution of amplitudes within a predetermined range. The obtained time series of flow velocity upstream of the flame and heat release rate fluctuations are used to train the nonlinear model using a multi-layer perceptron. Several models with varying hyperparameters are trained and the dropout strategy is used as a regularizer to avoid overfitting. The best performing model is subsequently used to compute the flame describing function (FDF) using mono-frequent excitations. In addition to accurately predicting the FDF, the trained neural network model also captures the presence of higher harmonics in the flame response. As a result, when coupled with an acoustic solver, the obtained neural network model is better suited than a classical FDF model to predict limit cycle oscillations characterized by more than one frequency. The latter is demonstrated in the final part of the present study. We show that the RMS value of the predicted acoustic oscillations, together with the associated dominant frequencies are in excellent agreement with CFD reference data.

\* Corresponding author at: GE Aviation Digital, Freisinger Landstr. 50, Garching 85748, Germany.
 *E-mail address:* nilam.tathawadekar@ge.com (N. Tathawadekar).

## 1. Introduction

Stringent emission regulations promote the development of lean premixed combustion systems. In gas turbines, this development is hindered by the risk of thermoacoustic instability, which causes undesired vibrations that may severely damage the engine. Thermoacoustic instabilities arise from the positive interaction between acoustic waves and unsteady combustion, and manifest themselves as self-excited acoustic oscillations [1]. To predict the amplitude of these acoustic oscillations at the design phase, low-order methods have proven to be useful [2–4] if combined with appropriate models for the nonlinear acoustic flame response. A well-established strategy to model such a nonlinear flame response is by means of the so-called flame describing function (FDF). It characterizes the fluctuations of heat release rate, $\dot{q}'$, produced as a response to upstream velocity perturbations, $u'$, which are characterized by a well defined frequency $\omega$, and amplitude $|u'|$. It reads

$$FDF: F(\omega, |u'|) = \frac{\dot{q}'(\omega, |u'|)/\bar{\dot{q}}}{u'(\omega, |u'|)/\bar{u}} \quad (1)$$

where $^-$ and $'$ denote temporal averaging and fluctuations, respectively. Conventionally, to obtain this FDF, the inlet velocity is excited harmonically and both the frequency and amplitude of the harmonic are varied. Although useful, the FDF approach is restrictive as it assumes that the flame responds exclusively at the same frequency as the input signal $u'$. Accordingly, it neglects the possible apparition of harmonics in the response $\dot{q}'$. Such an assumption may lead to an inaccurate prediction of thermoacoustic limit cycles [5,6].

To extend this FDF framework, Haeringer et al. [7] proposed an extended FDF (xFDF), which includes additional transfer functions relating higher harmonics of the heat release rate to the forcing velocity. Alternatively, Orchini and Juniper [8] introduced the flame double input describing function, where the FDF is extended by forcing the flame with two amplitudes and two frequencies.

A more reliable yet more complex alternative to the FDF-acoustic network approach is the hybrid CFD/low-order acoustic approach. It consists in coupling 'on the fly' Computational Fluid Dynamics (CFD) to an acoustic solver. While the former simulates the flame acoustic response $\dot{q}'$ to a given input velocity $u'$, the latter provides the fluctuations $u'$ as the result of the acoustic source $\dot{q}'$ and the propagation of acoustic waves and corresponding reflection at the domain boundaries. This method was demonstrated to predict limit cycle amplitudes accurately [9,10]. Unlike FDF, this approach captures the nonlinear interaction between fundamental frequency and associated harmonics to predict the self-excited oscillations.

Obtaining the FDF (or xFDF) by means of CFD is usually computationally expensive. This is because, under traditional methods, many independent simulations are necessary, each one associated to a given amplitude and frequency. The computational costs for the double input FDF are even higher if the same strategy is followed.

The computational costs associated with the hybrid CFD/low-order model are directly linked to the costs of CFD, which may be considerable for highly resolved flames. As a consequence, nonlinear thermoacoustic studies, where the nonlinear flame response plays a central role, are generally unaffordable if all the calculation chain relies on numerical simulations. It is therefore crucial to develop methodologies, where the FDF – or eventually more complex flame models – can be obtained by a *single* CFD simulation, where a carefully designed input signal $u'$ is applied during a short physical time window. Such a method should be analogous to the CFD/SI method [11], where broadband excitation is applied to CFD and combined with system identification (SI) to capture a linear flame response, also called a flame transfer function (FTF), in *one* shot.

The present study does not use SI methods, as they are generally devoted to linear time invariant systems [11], but instead uses the data collected by broadband forcing and makes use of a machine learning based approach to model the nonlinear acoustic response of flames. Previous to the present work, Selimefendigil and co-workers [12,13] and Förner and Polifke [14] had already extended the CFD/SI approach to nonlinear regimes using neural networks in the context of nonreacting flows. Another approach was also proposed by Jaensch and Polifke [15] for reacting flows, where they attempted to model the FDF of the Kornilov's flame [16] using neural network models. The results of that study were not satisfactory and possible reasons for this will be discussed in Section 3.3. Nonetheless, it suggested different ways to improve the results, such as using a more sophisticated and robust model.

Following the work in [15], we formulate a regression problem, which is solved using a multilayer perceptron (MLP). The universal approximation theorem states that, an MLP has the ability to learn any nonlinear function in its subspace [17]. This forms the basis to use MLPs for the modeling of the FDF. The objective of this paper is twofold: (i) to recover the FDF of a laminar flame using a neural network (NN), and (ii) to predict the limit cycle amplitudes by coupling the trained NN model with an acoustic solver. The approach proposed here, which is based on neural networks, can be applied in both the linear (for the FTF) and nonlinear (for the FDF) regimes, using data of one unsteady CFD simulation. The paper is organized as follows. Section 2 describes the experimental and numerical setup of the Kornilov flame [10,16,18]. The neural network-based modeling methodology is then discussed in Section 3. The results of the trained NN to model the FDF and obtain the limit cycle amplitudes are shown in Section 4, where they are also compared with the previous work by Jaensch and Polifke [15]. A summary of the results and future work is presented in the final section.

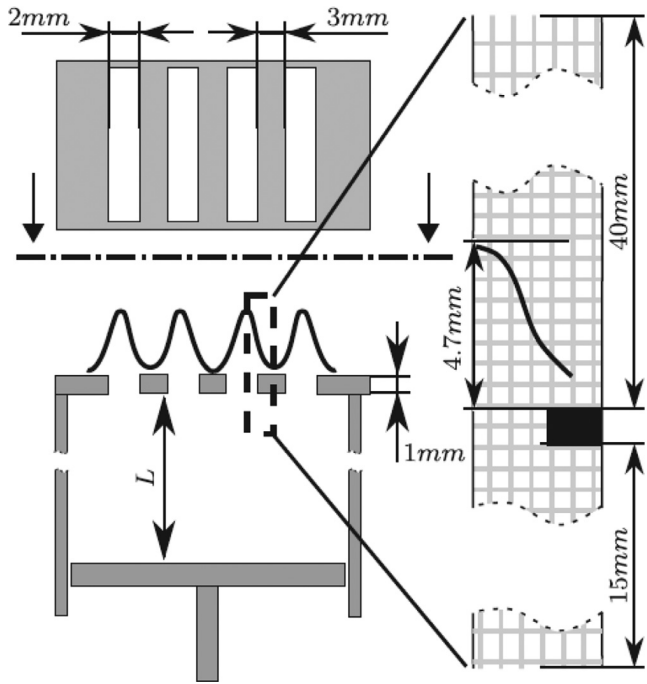

Fig. 1. Left: Experimental configuration. Right: CFD domain. Figure adapted from [10].

## 2. Experimental and numerical setup

The experimental and CFD setup studied in this work is shown in Fig. 1. It is the laminar multi-slit burner investigated by Kornilov et al. [16] where a premixed methane-air mixture with an equivalence ratio of 0.8 is used. The velocity perturbation is performed by using a loudspeaker operated by a single tone generator.

Here, the numerical setup and results of [15] are used: a 2D CFD domain with symmetric boundary conditions to reduce the computational cost, the two-step chemical scheme as detailed in [19] and OpenFOAM, a low-Mach number solver, are used for the simulation. Additionally, no combustion model is used here as the flame is laminar and all species transport equations are fully resolved. This approach with a low-Mach formulation, coupled to an acoustic network model via the global heat release rate and the fluctuation of the axial velocity at a reference position upstream of the flame, has already been validated against one based on resolving the fully compressible Navier–Stokes equations, coupled to the low-order model via the characteristic wave amplitudes at the inlet boundary by Jaensch et al. in [10]. Given that there is no turbulence, all flow quantities are also perfectly resolved. The grid was chosen following the grid independence study performed in [10]. At the inflow, a mean inlet velocity of 0.4 m/s and inlet temperature of 293 K are imposed. The plate on which the flame is stabilized is modeled as a no-slip wall with a fixed temperature of 373 K, as measured in the experiment [16]. A structured grid with 122,300 cells was used. The grid is uniform with a cell size of 0.025 mm in the region of the steady-state position of the flame, which ensures that the flame is fully resolved, and in the area of contractions. Outside this region, the cells were stretched in the axial direction [10]. The CFD simulation is run with an adaptive time-stepping scheme with an average timestep $\Delta t = 10^{-6}$.

The time series for the training of the NN used in this work are those obtained by Jaensch and Polifke [15]. The excitation signals used are of broadband nature as shown in Fig. 2. The signals allow for a broadband, low-pass filtered excitation spectrum with constant amplitude across the frequency range of interest [20]. A non-Gaussian simulation method is applied to generate the random signal time series based on a prescribed power spectrum and a cumulative distribution function [21]. $u'$ is the area averaged velocity measured upstream of the burner plate. It is normalized by extracting the temporal average. The normalized excitation amplitudes are 0.5, 1.0 and 1.5 for the three datasets. These are subsequently called datasets A, B and C, respectively. Despite having similar PDF of ampli-tudes, all three forcing signals are generated independently and are statistically independent and uncorrelated. The heat release rate fluctuations are integrated over the entire flame and then normalized to get the $\dot{q}'$ time-series. It should be noted that the accuracy of the NN model depends on the accuracy of the training data used. Hence, the quality of the CFD simulation needs to be ensured. In the present study, given the laminar state of the flame and the fact that the CFD resolves all thermochemical and flow time and lengthscales, the data is accurate enough to be used for the training of the NN model.

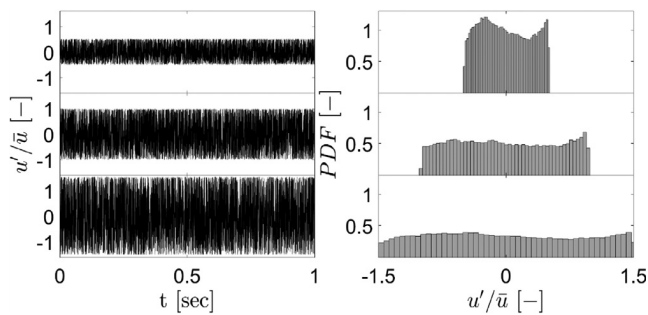

Fig. 2. Input signals used to train and validate the NN models. Datasets A, B, C from top to bottom. Left: input time series $u'/\bar{u}$, Right: probability density function (PDF) of the input data.

## 3. Neural network

In this section, we formalize the problem statement and provide the implementation details of the neural network model.

Given a time series data $\mathcal{D} = \{u'(t), \dot{q}'(t)\}_{t=0}^{m\Delta t}$, where $u'(t) \in \mathbb{R}$ and $\dot{q}'(t) \in \mathbb{R}$, the objective is for the neural network to find a mapping $\mathcal{F}$ such that

$$\dot{q}'(t) \approx \mathcal{F}(u'(t)|u'(t-\Delta t), \ldots, u'(t-n\Delta t)) \quad \forall t \tag{2}$$

where $\Delta t = 15\,\mu s$ denotes the sampling time. The main assumption is that the output at any point in time $t$, i.e. $\dot{q}'(t)$ depends not only on the input $u'$ at time $t$, but also on some history of input as formulated in Eq. (2), which reflects the time-lagged nature of the system. Here, we consider at least 10 ms of history as it is the characteristic time scale of the response of the system [22]. To solve this problem, a Multi-Layer Perceptron (MLP) is used and its architecture is described next.

### 3.1. Non-linearity through activation functions

A multi-layer perceptron is a feed-forward network mapping the input to the output as represented in Fig. 3 where an input layer, two hidden layers and an output layer are represented. Here, the input $\underline{u}' = [u'(t), u'(t-\Delta t), \ldots, u'(t-n\Delta t)]$ is mapped to the output $\dot{q}'(t)$ via the concatenation of nonlinear functions. The predicted output of the MLP is denoted by $\dot{q}'_p(t)$. It is termed as a feed-forward network as the output is not fed back into the input layer. Each hidden layer consists of neurons which are fully connected (FC). This means that each and every neuron in layer $l_{i-1}$ is connected to all neurons in layer $l_i$ through a weight matrix $W_i$. Hence, the intermediate output of the hidden layer $i$ can be written as $Z_i = W_i^T X_{i-1} + b_i$, where $X_{i-1}$ is the output of layer $i-1$ and $b_i$ is a bias term. In MLP, the non-linearity is introduced by using an activation function $g$ and, therefore, $X_i = g_i(Z_i)$ becomes the final output of layer $i$. The activation function is one of the most important choices in the design of MLP and the main options are usually the logistic function (sigmoid), hyperbolic tangent (tanh) or rectified linear unit (ReLU). Here, the tanh activation function is chosen because it generally allows for a faster convergence of the training of the NN compared to the sigmoid function [23]. Though ReLU is found to be better than tanh for many applications [24,25], functions approximated with tanh activation units are smoother than ReLU which makes tanh more appropriate for the present work. The tanh activation function is used for all hidden layers, whereas the linear activation function is adopted for the output layer.

### 3.2. Dropout as regularizer

The performance of many machine learning algorithms, neural networks included, suffers from overfitting. That is why it is necessary to regularize the model to reduce the risk of overfitting and improve the generalization error. One way to "regularize" a model is to combine several models with different architectures or different data [26]. However, it is computationally expensive to find optimal hyperparameters for each architecture and there may not be enough data available to train the network on different subsets of the data. The dropout regularizer addresses both of these issues. The key idea is to randomly drop units along with their input and output connections from the neural network during the training as shown in Fig. 3. This prevents the units from co-adapting too much [26]. The dropout rate is one of the design parameters and, in this work, it is set at 0.5 for the hidden layers. That means that neurons in the hidden layers are dropped with a probability of 0.5 during training. This was found to be an efficient value for this work.

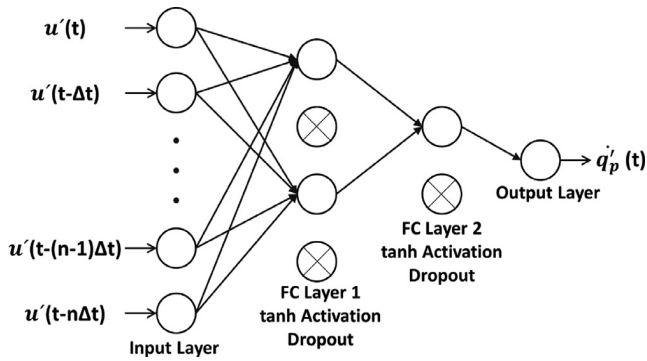

Fig. 3. Typical representation of a neural network model with two hidden layers and dropout regularizer.

### 3.3. Neural network architecture

To train the neural network, a loss function based on the mean squared error (MSE) is used:

$$MSE = \frac{1}{n}\sum_{i=1}^{n}\left(\dot{q}'_p(i) - \dot{q}'(i)\right)^2 \quad (3)$$

where $\dot{q}'_p$ is the prediction from the neural network. The network parameters (weight matrix $W$ and bias $b$) are updated in response to the loss function using the Adam optimizer with the learning rate as one of the hyperparameters.

In total, in order to get optimal performance, the design of the MLP requires choosing the values of these hyperparameters: the number of hidden layers, number of hidden units and the learning rate of the optimizer. Generally, a grid search is widely used for hyperparameter optimization but it is computationally expensive. So, here, a random search is used as it is found to be efficient and inexpensive compared to grid search [27]. Neural networks with up to 4 hidden layers and a maximum of 200 neurons in the first hidden layer were explored and the learning rate was log-linearly sampled from the range [0.0001, 0.001].

The present study uses python – TensorFlow (http://tensorflow.org/) as a framework. Contrary to the modeling approach of Jaensch and Polifke [15], where they use the default implementation of artificial neural networks in Matlab R2015b. Additionally, we explore here deeper (more no. of hidden layers) and wider (more no. of neurons per layer) networks compared to [15]. Regularization is also found to be crucial to avoid over-fitting. We use dropout as a regularizer, whereas [15] did not use any form of regularization. These differences between our work and [15] may explain the significant differences in the corresponding results shown next.

## 4. Results

### 4.1. Neural network model

Figure 2 shows the three datasets used to train and test the neural network models. Each dataset contains 1 s of inlet flow velocity fluctuations and associated heat release rate fluctuations data. For the training and testing, each dataset is divided into 0.7 s for training and 0.3 s for testing, where the last 20% of the training data is used for validation. As described in Section 3, the input data is pre-processed to include the history of inputs. This transforms the input data into a higher dimensional space and is fed to the neural network in batches. The model with the smallest MSE on the validation set is taken to be the best and its performance is evaluated against the test set. For the three datasets, the models are trained for different hyperparameters settings and the best performing model is stored for further use. Table 1 shows the number of hidden layers, neurons per layer and number of trainable parameters (weights) for the best performing neural network. A dropout layer is used between each hidden layers. The last layer of the network is the output layer with 1 neuron. Figure 4 shows the prediction of the best trained model against the reference from CFD on test data for all three datasets. This figure highlights the good agreement between the NN model and the CFD on the dataset available.

### 4.2. Forced response

To further assess the trained neural networks, the FDF is obtained from them and compared to CFD results. To do so, the trained neural networks are forced with mono-frequent excitation of different amplitudes. Frequencies from 10 to 500 Hz and amplitude levels of 0.5, 1.0 and 1.5 are considered. Figure 5 shows the comparison of the FDF computed from the best neural network models and the CFD simulation. It is seen that all the NN models capture the FDF at amplitude = 0.5. However, only the NN trained on dataset C captures the FDF accurately for all amplitudes. The networks trained on datasets A and B fail to capture the higher amplitude flame response because those datasets lack the necessary amplitude information as shown in Fig. 2. For dataset C, all amplitude levels are present and quite uniformly distributed. Therefore, that dataset contains all the necessary information that the NN needs to learn the non-linear flame response. However, as the phase does not vary much with different amplitude levels, the networks trained on all datasets are able to predict the phase of the FDF satisfactorily. The shaded regions in Fig. 5 show the bounds of prediction, using the estimates of the 5 best performing NNs. The dataset C exhibits tighter bounds, which indicates less uncertainty around the prediction.

Table 1
Details of the best performing MLP for all datasets.

| Dataset | # Hidden layers | # Neurons per layer | Total weights |
| --- | --- | --- | --- |
| A | 2 | [73 36] | 75,774 |
| B | 2 | [92 46] | 96,417 |
| C | 3 | [128 64 32] | 138,497 |

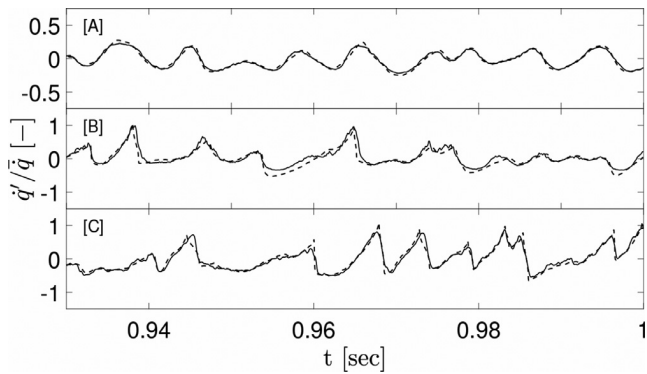

Fig. 4. Validation of the forced response of the best fit neural network model against broadband time series for datasets A, B, C from top to bottom. Solid line: prediction of heat release rate fluctuations by NN model on test data. Dashed line: CFD reference.

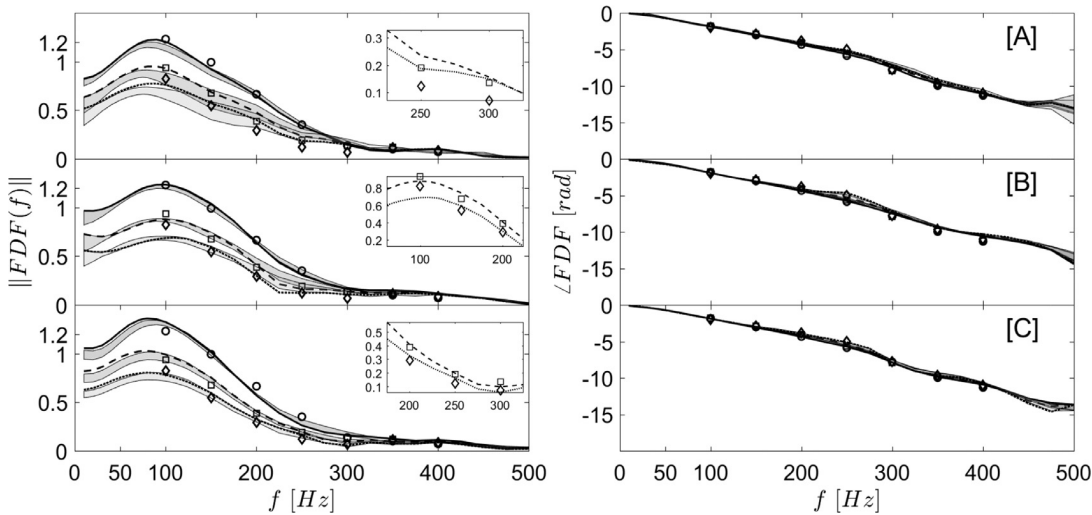

Fig. 5. Comparison of the FDF computed from the best NN model and from the CFD simulation for datasets A, B, C from top to bottom. Left: Gain; Right: Phase; Lines: estimate by the best NN fit model. Markers: CFD reference using mono-frequent excitation. Shaded area: Bounds of prediction by top 5 NN models. Excitation amplitudes: 0.5 (solid line, circle), 1.0 (dashed line, square), 1.5 (dotted line, diamond).

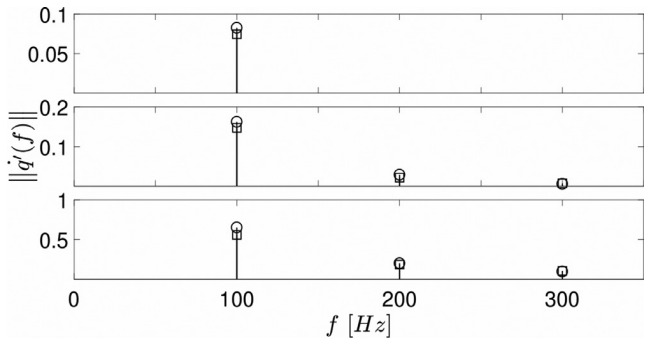

Fig. 6. Comparison of FFT component of heat release rate fluctuations at different harmonics for an excitation signal with frequency = 100 Hz and amplitudes from top to bottom 0.05, 0.1, 0.5. Circle: NN model. Square: CFD reference.

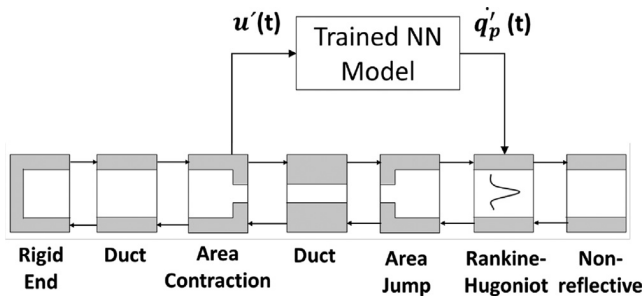

Fig. 7. Integration of acoustic network and neural network models.

As discussed in the introduction, the FDF neglects the excitation of higher harmonics and the ability of the NN to do this is also assessed here. The trained NN-model is excited with 3 different sinusoidal signals: each with frequency of 100 Hz and amplitudes of 0.05, 0.1 and 0.5, respectively, and the results are compared with the equivalent excited CFD simulation. Figure 6 shows the FFT contents of the heat release rate fluctuations at the fundamental frequency and its higher harmonics. At 0.05, the output is purely sinusoidal. For amplitudes of 0.1 and 0.5, higher harmonics start to appear which is a trend that the NN-model also predicts. This behavior is consistent across all other frequencies. This shows that the proposed neural network is capable of learning more complex interactions between harmonics than the FDF.

### 4.3. Coupling of neural network and acoustic network

From the previous sections, it was observed that only the neural network trained with dataset C can appropriately reproduce the FDF for all frequencies and amplitude levels under consideration. The lack of accuracy of the NNs trained with datasets A and B might lead to an inaccurate evaluation of the acoustic limit cycle. Therefore, for the rest of the analysis, only neural network model trained on dataset C will be used for further processing and to assess whether it can accurately predict self-excited oscillations. To do this, the trained neural network model is integrated into an acoustic network model through flow velocity fluctuations and heat release rate fluctuations [28], as shown in Fig. 7, where the overall network represents a flame in a duct. Additional details about the acoustic network model can be found in Emmert et al. [29].

In this section, the results of the coupled neural network and acoustic network models are compared with the hybrid CFD/low-order model by Jaensch et al. [10]. In that work, the flame response was obtained using CFD. To study the stability of the system, the length of the plenum duct is varied from 50 mm to 1000 mm and the amplitudes of the velocity fluctuations are collected. Figure 8 shows the RMS value of the flow velocity fluctuations measured at the upstream side of the burner plate and the dominant frequency of the self-excited oscillations for different plenum lengths. It is observed that the system becomes unstable at plenum length of 160 mm which is in agreement with the CFD results from [10]. Additionally, the neural network model shows a good agreement with the CFD results with a slight over prediction for longer plenum. Furthermore, the NN model also exhibits the same trend as the CFD results where, initially, the second harmonic is dominant and, as the plenum length increases, the fundamental frequency becomes dominant (see Fig. 8 at frequencies for plenum lengths between 200 mm and 300 mm). Finally, the shape and frequency response of the obtained self-excited oscillations for two different plenum lengths are shown in Fig. 9. The snapshots of the time series are taken once the steady limit-cycle state is reached. An excellent agreement is observed between the neural network based approach and the CFD for both plenum lengths, with the appropriate frequency being dominant in the response signals. This shows that the NN-based approach can provide additional information compared to the FDF method as it also produces harmonics in its response.

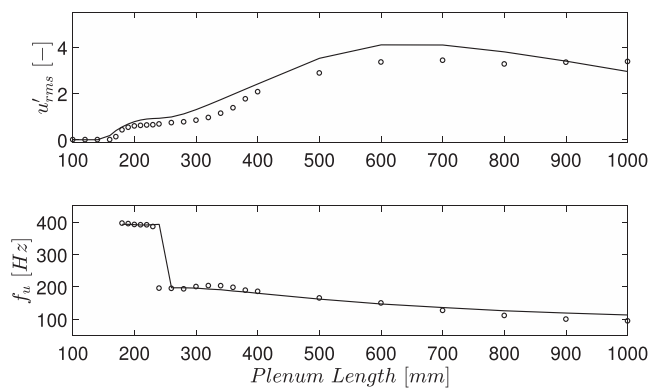

Fig. 8. Comparison of RMS and dominant frequency of the reference velocity for different plenum lengths. Line: estimate by the NN model. Markers: CFD reference.

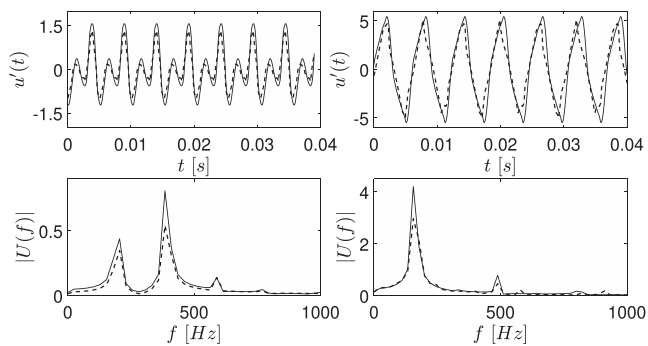

Fig. 9. Time series (Top), power spectrum (Bottom) of the velocity signal for neural network model integrated with an acoustic network (NN model) and the hybrid/low-order model by Jaensch et al. (CFD reference) for two plenum lengths ($L$). Left: $L = 200$ mm, Right: $L = 500$ mm. Solid Line: NN model, Dashed Line: CFD reference.

## 5. Conclusion

Multi-layer perceptrons with different numbers of hidden layers and variable neurons were explored to model the nonlinear flame response of a Bunsen-type flame. The neural network models showcase the ability to learn the FDF for the laminar premixed flame under study, while using only one CFD simulation. The simulation is excited with a broadband signal characterized by a uniform distribution of all amplitudes of interest. The trained neural network model captures the flame response not only at the input frequency but also at higher harmonics. When coupled with an acoustic solver, the trained NN could accurately reproduce the bifurcation diagram and the limit-cycle amplitude, shape and frequency contents when compared to the hybrid CFD/acoustic solver of Jaensch et al. [10].

The present work demonstrates the use of NN for CFD/SI approach in single-input, single-output systems. To extend this approach to multi-input, single or multi-output systems, multiple CFD simulations could potentially be required to capture the variance in multiple input systems and the inputs to the NN need to be modified to account for multiple lags associated with such systems. These modifications will be explored in future studies. Another important direction of future work includes determining the minimum signal time length required to have an acceptable accuracy of the trained network. Additionally, future work will be dedicated to assessing the proposed approach to a turbulent flame. In that case, regularization, introduced here via the dropout layer, will play a crucial role in the flame response evaluation, as strong levels of noise may be present.

**Declaration of Competing Interest**

The authors declare that they have no known competing financial interests or personal relationships that could have appeared to influence the work reported in this paper.

**Acknowledgments**

N. Tathawadekar acknowledges the financial support of the European Union's Horizon 2020 research and innovation program under the Marie Sklodowska-Curie grant agreement No. 766264. N.A.K. Doan acknowledges the support of the Technical University of Munich – Institute for Advanced Study, funded by the German Excellence Initiative and the European Union Seventh Framework Programme under grant agreement no. 291763. The authors would like to thank Prof. Polifke for his comments on the draft version of this paper, and Dr. Stefan Jaensch for providing the simulation data.